\documentstyle[aasms,12pt,epsf]{article}
\def \aa    #1 #2   {{ A\&A \/}, {#1}, {#2}}
\def \aas   #1 #2   {{ A\&AS \/}, {#1}, {#2}}
\def \aj    #1 #2   {{ AJ \/}, {#1}, {#2}}
\def \apj   #1 #2   {{ ApJ \/}, {#1}, {#2}}
\def \apjs  #1 #2   {{ ApJS \/}, {#1}, {#2}}
\def \mnras #1 #2   {{ MNRAS \/}, {#1}, {#2}}
\def \prl   #1 #2   {{ Phys. Rev. Lett. \/}, {#1}, {#2}}
\def \nat   #1 #2   {{ Nature \/}, {#1}, {#2}}
\def \com   #1 #2   {{ Comments Astrophys.\/}, {#1}, {#2}}
\def \sast  #1 #2   {{ Soviet Astron. \/} {#1}, {#2}}
\def \sastl #1 #2   {{ Soviet Astron. Lett.\/}, {#1}, {#2}}
\def \astr  #1 #2   {{ Astrophysics \/}, {#1}, {#2}}
\begin{document}
\baselineskip 0.8cm
\title{UNIVERSALITY OF THE
NETWORK AND BUBBLE TOPOLOGY IN COSMOLOGICAL GRAVITATIONAL SIMULATIONS}

\author{Capp Yess and Sergei F. Shandarin \\
        University of Kansas, Lawrence, KS 66045}
\begin{abstract}

Using percolation statistics we, for the first time, demonstrate the
{\it universal} character of a {\it network} pattern in the real space, mass
distributions
resulting from nonlinear gravitational instability of initial
Gaussian fluctuations. Percolation analysis of five stages of the nonlinear
evolution of five power law models ($P(k) \propto k^n$ with $n=+3,~+1,~0,~-1$,
and $-2$ in an $\Omega =1$ universe) reveals that all models
show a shift toward a network topology
if seen with high enough resolution. However, quantitatively,
the shift is significantly different in different models: the smaller the
spectral index $n$ the stronger the shift.
On the contrary,
the shift toward the {\it bubble} topology is characteristic only for the
$n\le-1$ models.
We find that the mean density of the percolating structures in the nonlinear
density distributions generally is very
different from the density threshold used to identify them and corresponds
much better to a visual impression. We also find that the maximum of the
number of structures
(connected regions above or below a specified density threshold)
in the evolved, nonlinear distributions
is always smaller than in Gaussian fields with the same spectrum,
and is determined by the effective slope at the cutoff frequency.

{\it Subject headings}: cosmology-galaxies:clustering-methods:numerical

\end{abstract}

\newpage
\section{INTRODUCTION}

The topology of the galaxy distribution can provide important clues to the
formation of large-scale structure in the universe and the nature
of the initial density fluctuations (\cite{Sh-Z83}).
The recent compilation of large galaxy surveys, large in galaxy numbers
and survey volumes, has prompted a new round of topological studies of
the structure of the universe.  The amount of information in studies like
the IRAS and CfA Redshift Surveys, and the upcoming Sloan Digital Survey
suggests that the major problems of discreteness, boundary effects, and local
inhomogeneity that have plagued topological analysis to date may be overcome.

Two types of methods have been developed and employed to quantitatively assess
the topology of the galaxy distributions of these surveys.
The first one is based on the evaluation of the mean Euler characteristic,
$\chi(\nu)$, or the genus curve, $g(\nu)$,
as functions of the density contrast: $\nu = \delta/\sigma_{\delta}$
($g=-\chi/2$) (\cite{Dor70};
\cite{Bar-etal86}; \cite{Got-Mel-Dic86}) \footnote{For a general review of
this method see e.g. (\cite{Mel90})}.
The genus curve method has been utilized for studies of many galaxy
catalogs (see e.g., Weinberg, Gott \& Melott 1987; Moore et al. 1992;
Vogeley et al. 1994).

The other method is based on percolation theory.
Percolation is the study of the number and various properties of
``clusters''. \footnote{In the absence
of a better term, we label
as ``clusters'' the high density regions bounded
by the surfaces of chosen constant density and ``voids'' as bounded regions of
low density. These terms are not to be confused with the astronomical terms,
voids and clusters of galaxies.}
In 1982, Zel'dovich noticed that the percolation properties of the
nonlinear density distribution in the HDM
(Hot Dark Matter) model are very different from that in the initial
Gaussian field.
In particular, the formation of a percolating cluster (the cluster spanning
through the entire system) happens more effectively
than that in Gaussian fields.
He also suggested characterizing the topology of nonlinear density
distributions by their percolation thresholds (\cite{Z82}).
Following Zel'dovich's
idea, one of the authors of this paper (S.Sh.)
suggested using percolation properties
of {\it the galaxy distribution} as an objective quantitative measure of the
topology of the large-scale structure and also as a discriminator
between cosmological models (Shandarin 1983; Shandarin \& Zel'dovich 1983).

A percolation technique was utilized in the study of the CfA I catalog
(\cite{Ein-Kly-Saa-Sh84}) and showed that the large-scale
distribution
of galaxies had a network structure. Theoretical studies of models
with a power law initial spectra showed that the $n=-1$ model clearly
percolated better than the $n=0$ model, and in the  $\Omega=1$ universe,
the $n=-1$ model was in agreement with observations (\cite{Bha-Bar83}).
The percolation
method also showed that the CDM (Cold Dark Matter) model appeared to have
a connected
rather than clumpy density distribution (Melott et al. 1983; Davis et al.
1985).
Later it was pointed out that the major disadvantage of any percolation
technique was the dependence
of the percolation thresholds on the mean density of the sample
(\cite{Dek-Wes85}) which made it difficult to calibrate for  sparse samples.
Similarly, we note that at
present some believe that sparse samples can only be reliably used for
the estimation of the two-point correlation function (\cite{Bou95}).
We elaborate on this question below.

For continuous fields such as density
fields the independent parameter is the density threshold.  A cubic lattice is
superimposed on the field, and lattice cells with densities above the
threshold are tagged as over-dense while cells below the threshold are labeled
under-dense. Over-dense cells are considered clusters, and can merge with other
over-dense cells to become larger clusters by satisfying a nearest neighbor
condition.  The nearest neighbor criterion we employ is that over-dense cells
must share a common side.  On a simple cubic lattice, this means that each cell
can have up to six nearest neighbors \footnote{For the versions
of percolation analysis
utilizing other nearest neighbor definitions or lattice structures see for
example, Mo \& B$\ddot{o}$rner 1990; de Lapparent, Geller \& Huchra, 1991.}.
As the
value of the threshold density is decreased, more cells will be tagged as
over-dense, and clusters will grow in number and/or size.  This process
proceeds until the largest cluster spans the available space and percolation
is said to have occured.  Void percolation
is accomplished in an analogous fashion except that the density threshold
is initially set at a low value, and voids grow as the threshold value is
increased.  To quantify the study of the clusters (voids) formed by the
above scheme, we trace the value of the three most robust parameters:
the cumulative
distribution function, referred to as the filling factor; the volume of the
largest structure (cluster or void) as a fraction of the corresponding
filling factor; and
the total number of all clusters and voids as a function of the
filling factor.  The
rationale for using the filling factor instead of the density contrast as a
fundamental parameter is that the filling factor is normalized allowing a
direct comparison of the parametric values of Gaussian and nonlinear
distributions as well as the topology of over-dense and under-dense phases.
Also we wish to isolate the information stored in the dependence of
the filling factor on the density threshold
from that stored in the other statistics.

The percolation threshold of a Gaussian distribution, as described
above, is believed to coincide with a change in the sign of the genus;
however, there is no theorem proving this.
Intuitively, it is plausible
that percolation thresholds approximately coincide with changes of the genus
sign. Sathyaprakash, Sahni \& Shandarin (1995) showed that in the nonlinear
distributions both transitions happen at close but significantly different
filling factors.

One advantage of the genus method is the existence of an analytic expression
for Gaussian random fields (\cite{Dor70}).
Tomita (1986) gave a very elegant analytic expression for the mean Euler
characteristic for multi-dimensional Gaussian fields, and
recently an analytic expression was obtained in the weakly
nonlinear regime (\cite{Mat94}).
However, one should not forget that the mean genus is a statistical
measure and therefore an estimate of the dispersion is needed before it becomes
meaningful. The dispersion of the genus for finite samples having
finite resolution has not been obtained analytically and
can be estimated only from numerical simulations. Percolation parameters
are also calculated numerically, but if the dispersion can be estimated,
the mean can also be estimated with similar accuracy.

It has been claimed that the percolation thresholds are the most sensitive
discriminators of varying models (\cite{Sh83}). The recent study of the
CfA II catalog using the genus method (\cite{Vog-etal94})
supports that suggestion.  The authors reduced the
information of the genus curve to three numbers one of which was
the genus peak width, $W_{\nu}=\nu_+ -\nu_-$ (where $\nu_+$ and
$\nu_-$ are the levels at which the genus changes sign).
Figures 12 through 14 in Vogeley et al. (1994) clearly
demonstrate that $W_{\nu}$ has the highest discriminating power among three
suggested parameters.
However,
we still believe that the percolation thresholds as well as $\nu_+$ and
$\nu_-$ should be interpreted separately because
they carry independent information about the topology of the structure.

The major improvement of the percolation technique we report in this
paper is related to the development
of an extremely efficient code for finding all the clusters at a given density
threshold (\cite{Sta-Aha92}; \cite{Kly-Sh93}). This enables us to calculate
more parameters with finer density thresholds variations than in earlier
studies. In this paper, we shall
report the results of studying the nonlinear, density distributions
in real space obtained from N-body simulations of the scale free cosmological
models: $P(k) \propto k^n$ with $n=+3,~+1,~0,~-1,$ and $-2$ in the $\Omega=1$
universe.
The volume of the largest structure will be used in this paper to indicate
the onset of percolation and characterize the topology of the distribution.
Every generic density distribution at a sufficiently high density threshold
will look like a system of isolated (non-percolating) clusters. At some lower
threshold the connected system spanning through the entire volume will
inevitably form. One possible definition of a network distribution would be
a distribution for which percolation occurs at a specified density threshold.
In this case the choice of the threshold must be physically justified.
We do not know how to select this particular density threshold. Therefore
we adopt a different definition of a network distribution.
Fields that percolate at lower filling factors than Gaussian fields (by
definition structureless) will be interpreted as having a ``connected''
or network structure; while fields that percolate at higher filling
factors will be considered ``disconnected'' or clumpy. Similarly we label
a distribution as having a bubble topology if the under-dense region
percolates at a higher filling factor than in a Gaussian field.

It is worth stressing that the terminology in this field is often confusing.
We use the labels clumpy, network, and bubble structure to emphasize the
degree of connectedness only.
In this paper we ignore the geometrical aspect of the
problem. For instance, we do not distinguish between a network made of
filaments (quasi one-dimensional objects) or pancakes or sheets
(quasi two-dimensional objects). In realistic distributions, it is
often impossible to
rationally assign a label to the shape of a particular density enhancement.
Various shape
statistics (\cite{Vis86}; \cite{Bab-Sta92}; \cite{Luo-Vis95}) assume that
the shapes can be approximated by a triaxial ellipsoid. However, the dynamics
of the nonlinear gravitational evolution suggests that the first collapsed
objects (pancakes) have a bowl like shape (\cite{Arn-Sh-Z82}; \cite{Sh-etal95})
which is very poorly approximated by a triaxial ellipsoid. In particular, the
thickness of such a structure would be highly exaggerated if it is approximated
by an ellipsoid.  This problem is addressed in an upcoming paper by
Sathyaprakash, Sahni \& Shandarin (1995).

In addition to determining the topology of the density distributions,
we will demonstrate a method of estimating the slope of the power
spectrum characterizing a distribution.
The  maximum of the number of clusters (voids) statistic is determined
by the effective spectral
index at the Nyquist frequency (or the smoothing scale). Potentially,
this relation can be used for measuring the slope of the spectrum.

In \S 2 we will explain the parameters we use to characterize the density
distribution in space and discuss their application to
the power law Gaussian fields. In \S 3 we describe the percolation of Gaussian
fields, and in \S 4 we describe the N-body simulations forming
the basis for our standardizations.  We detail the results from studying the
growth of the largest structures and elaborate on the number of
clusters results in  \S 5.  We summarize our findings in \S 6.

\section{METHOD}

\subsection{Filling Factor}

Percolation theory deals with the number and properties of the clusters.
The density threshold,
$\delta_c$, separating
over-dense ($\delta>\delta_c$) and under-dense ($\delta<\delta_c$) regions is
assumed to be a free parameter. When analyzing discrete distributions
(e.g. galaxy distributions), we assume a smoothing procedure
creating a continuous density distribution.

As we mentioned, the density threshold is not a convenient parameter
if linear (Gaussian) and nonlinear density distributions
or over-dense and under-dense phases are to be compared.
Instead we utilize the filling factor to parameterize the density threshold.
In this case one can easily compare the properties of clusters with those
of voids and also linear and nonlinear density distributions.
In a two-dimensional illustration it would be similar
to comparing different patterns provided
that the same amount of paint was used to make each pattern. For reference we
provide the relationship between the filling factor and the density contrast,
for the models studied (Figure 1).  The effect of evolution on the
relationship is evident in the graphs and underscores the reasons to use the
filling factor as a means of comparison. One sigma error bars are comparable
with or smaller than the thickness of the lines and are not shown.
Solid lines and dashed lines show the filling factor
for the nonlinear distributions and Gaussian fields with corresponding
spectra (see section 2.4).

We study various characteristics of a field as a function of the
filling factor which is the total volume occupied by the regions
having a density above (if we study over-dense regions) or below
(if we study under-dense regions) a specified threshold. The filling
factor coincides with the ``volume fraction'' used by Vogeley et al (1994).
However, instead of expressing it in terms of $\nu$ (the number of standard
deviations above or below the mean density of the isodensity contour filling
the same volume in a Gaussian field), we use it directly. The filling factor
is obviously related to the cumulative distribution function of the
corresponding phase:
\begin{equation}
f\!f(\nu) = {1 \over \sqrt{2\pi}}~\int_{\nu}^\infty e^{-t^2/2}~ dt.
\end{equation}
Also, in Gaussian fields there is no statistical difference between under-dense
and over-dense regions; therefore, the filling factor is the same for both.

\subsection{Largest Structures}

Here for definiteness we shall talk about over-dense regions, the under-dense
regions can be discussed similarly.
For a given density threshold we have a set of regions that differ by volumes,
shapes and other parameters. We select the one having the largest volume and
call it the largest cluster.
It is convenient to measure the largest cluster and void
as a fraction of
the corresponding filling factor. Thus, if the largest cluster is $0.9$ at
filling factor of $0.2$ it means that the density is higher than the chosen
threshold in $20\%$ of the volume and almost all of that volume
($90\%$) is comprised of only one connected region. The absolute volume
of the largest cluster is clearly $18 \%$ of the total volume of
the sample.

In a finite system, (like the density
distribution in the N-body simulations or galaxy surveys)
if we start from a very
high density threshold we do not find any clusters at all.
Lowering the threshold we find the first cluster corresponding to
the highest density peak.
Obviously, it is also the largest cluster,
and its volume measured in terms of the filling factor is unity. This is
an effect of a finite system and we shall ignore it.
In other words one cannot use the statistics at density thresholds which are
too high, or filling factors which are too small.
At lower thresholds we typically have many clusters. Usually
the volume of the largest cluster is negligible compared to the total
volume of all clusters (the filling factor). The actual volume of the
largest structure (in the units of the filling factor) can be used
as a measure of the fairness of the sample: the smaller the largest
structure at low values of the filling factor the better the sample.
Decreasing the density threshold we reach the situation where clusters
begin to merge.  As a result the largest cluster becomes a significant
fraction of the filling factor, and eventually almost all over-dense
regions combine to make just one cluster. At this stage the volume of the
largest cluster almost equals the
filling factor, and percolation is said to happen. In the past much
effort was spent on an accurate measurement of the percolation threshold
(\cite{Kly87}; \cite{Dom-Sh91}; \cite{Kly-Sh93}). In contrast, we use the
{\it largest structure as a function of the filling factor} as a practical
indicator of the percolation transition as suggested by Shandarin (1994, 1995).

The percolation transition is a universal behavior of every
non-degenerate system; however, the
filling factor at which the transition happens is particular to each system
and may vary. The percolation threshold marks a change in the topology of a
distribution. Above the percolation threshold ($\delta_c \ge \delta_p$)
the over-dense system consists of isolated clusters having finite
volumes, and the topology is of a meat-ball or clumpy type (if the under-dense
regions are considered it is usually called a bubble topology).
Bellow the percolation threshold ($\delta_c \le \delta_p$) most of the
over-dense volume is in one cluster spanning the entire system, and the
topology is labeled a network or sponge topology. Both terms, network
topology and sponge topology, correspond to the positive genus, but the term
network structure suggests that the percolating structure is ``thin'' which is
a geometrical rather than topological characteristic.

We plot the largest cluster and the largest void volume fractions
versus the corresponding filling factor in the same diagram. As we mentioned
before,  Gaussian fields exhibit no statistical difference between over-dense
and under-dense phases and both structures have similar properties. In
contrast, for all nonlinear density distributions the largest cluster
percolates as easy or easier
(that is at smaller filling factors) than the largest void which is
a manifestation of non-Gaussianity resulting from nonlinear
gravitational clustering.

\subsection{Number of Clusters and Voids}

The third type of statistic we use in this paper is the total number
of clusters and voids at a given filling factor. The significance
of this statistic is its sensitivity to the slope of the spectrum.
We shall measure the number of structures per Nyquist volume: $V_{Ny} =
\lambda_{Ny}^3 = 8$ mesh cells. The number of clusters and/or
voids typically grows with the growth of the filling factor until it riches
the value about 0.1-0.13 then it quickly decreases because of merging of the
clusters.
In the nonlinear distributions the number of clusters is often (but not always)
smaller than that of voids at the
same filling factors. Again the difference between the number of the
clusters and voids is a manifestation of non-Gaussianity of the
density field.

\subsection{Mixing Phases}

We produce three distinct types of density fields for analysis in this study:
random Gaussian fields with pure power law spectrum, fields derived from
N-body simulations with evolved nonlinear power spectrum, and Gaussian fields
with nonlinear power spectrum produced by randomizing the phases of the N-body
simulation distributions.  Percolation of Gaussian fields (with linear and
nonlinear power spectrum) produces standards for characterizing the
topologies and estimating the spectral indices of density fields in subsequent
studies.  The analysis of fields generated from N-body simulations checks and
calibrates our ability to describe fields by the method outlined above.

In the comparison of simulation fields to Gaussian fields, we wish to avoid
grid effects as much
as possible.  Because grid effects are inherent in the method, our solution is
to generate nonlinear Gaussian fields from the simulation fields, thereby,
assuring grid effects similar to those
displayed in the nonlinear, parent distributions. This also resolves
the question of how to calibrate the percolation curves for sparse samples.
A good way to do this is to start by Fourier transforming a
nonlinear N-body simulation field to $k$-space.  Then, in the resulting
transform generate random phases keeping the amplitude of every
wave exactly the same as before, and finally make an inverse Fourier transform
resulting in a new Gaussian field having the same power spectrum as the
original
nonlinear parent field. We ignore the fact that this field may have negative
values since we study only the percolation properties parameterized by the
filling factor. Since the generated Gaussian field has the same amplitudes
it must be affected by the finite resolution similarly to the nonlinear density
fields.  This randomized version of the simulation density field
is percolated to produce an additional standard for distinguishing the
topology of the model and to illustrate the relationship between phase
correlations and nonlinearity.
The comparison of the nonlinear and randomized (Gaussian)
fields provides a measure of the phase correlation in the nonlinear fields.
We show and interpret the collective
results of the percolation of Gaussian and simulation fields in the
following sections.

\section{PERCOLATION IN GRID GAUSSIAN FIELDS}

The ability to recognize Gaussianity in a distribution is important for two
reasons.  First random Gaussian fields are used as benchmarks for describing
the topology of density fields derived from survey data, and secondly, it
has been demonstrated that random phase density fluctuations could be produced
by an inflationary period in the early history of the universe
(see e.g. Linde 1992). Importantly, a random Gaussian field is fully
distinguished by its associated power spectrum.  If gravitational instabilities
were the only mechanisms of structure formation and the initial density
perturbations were Gaussian, then the initial power spectrum would determine
the large scale structure of the universe.

It is worth emphasizing that random fields generated
numerically on a grid are only approximately Gaussian. We shall call them
grid Gaussian fields. One obvious deficiency of a grid Gaussian field
compared to a true Gaussian field is that
the probability distribution function is always wrong for sufficiently
large values of the field. We will report another less obvious deficiency
related to the topology of the field.

The Gaussian fields we employ in this study are created by transforming an
array of coefficients, that are random Gaussian distributed, from $k$-space to
real space.  The range of $k$ was limited by both the box size
(the fundamental mode, $k_f$) and the resolution of the mesh
(the Nyquist wavenumber, $k_{Ny}$) such  that $k_f \le k \le k_{Ny}$.
We also study fields with cutoffs for which $P(k)= 0$ for
$k > k_c$.  The results are Gaussian random density fluctuation fields which
are homogeneous and isotropic.  These properties insure that the associated
power spectrum is a function of $k$ only.  The spatial modes of these
fields are
mutually independent and have random phases.  (We exploit this random phase
condition of Gaussian fields in our phase mixing of N-body simulations
explained above.)  It is important to remember that Gaussian fields are by
definition structureless.

Figure 2 shows several parameters calculated for grid Gaussian fields
with power law spectra: $P(k) \propto k^n$ with the spectral index
$n=-2,~-1,~0,~+1$.
The fields were generated on a $64^3$ mesh by
Fourier transform of random numbers, so no particles were involved.
The results for two cutoff values are shown: $k_c=16$, and $k_c=32$ (the
Nyquist wavenumber).  Four realizations with different random number seeds
were generated for each curve, and mean values are plotted. One  $\sigma$
error bars are also shown, but in some cases they are too small to be seen.

The two top panels show the largest cluster measured as
a fraction of the filling factor.
The left hand side panel shows the grid Gaussian functions
with the power law spectra cut off at $k_c=16$, and the right hand side
panel shows those without an arbitrary cutoff. According to percolation
theory, percolation in Gaussian fields occurs at $f\!f \approx 0.16$
(corresponding to $\nu = 1$) independently of
the power spectrum. The onset of percolation is marked by a sharp growth of
the largest cluster and corresponds to the change of the topology from
meat-ball to sponge type. The left top panel is in agreement with this
result, but the right top panel is not.
The right panel shows that the percolation
threshold depends on the spectral index; the larger the index the higher
the percolation threshold. Statistically the difference is highly significant.
Thus, the limited resolution of the mesh results in grid Gaussian fields, with
no spectral cutoff, having a meat-ball
topology shift indicated by a percolation
transition at greater filling factors than true Gaussian fields.
This may affect the appearance of the large-scale structure
in the N-body simulations especially in those having a small number of
particles, making them look more clumpy than they should.

At small filling factors the largest structure must be negligible in
sufficiently large samples. The finite size of the largest structure
can be used as an internal characteristic of the fairness of a sample.
The smaller the largest cluster in the non-percolating regime the better
the sample.
One can see that the more negative the spectral index the more difficult
it is to satisfy this condition.

The middle panels show the number of clusters per  Nyquist volume (8 mesh
cells) as a function of the filling factor. The curves reach their maximum at
filling factors of about 0.11 (dependent slightly on the spectral index $n$)
for the spectra with no arbitrary cutoff ($k_c=k_{Ny}=32$)
and drop with the growth of the filling factor because of the merging of
clusters. In the $k_c=16$ models the maxima are reached at filling factors
($f\!f_{max} = 0.08$) which are roughly half the value of the filling
factors at the percolation threshold regardless of the spectral index.
The value of the maximum (marked by a dot) depends very weakly on
the cutoff and is determined only by the spectral index
on the scale of the cutoff.
This demonstrates that  percolation analysis can discriminate between random
Gaussian fields typified by different spectral indices (for pure power law
initial power spectra) as easily as the maximum of the genus curve.

The left bottom panel shows the cumulative number density of peaks
calculated analytically using the equation
\begin{equation}
n_{pk}(f\!f) = ~\int_{-{\infty}}^{\nu} N_{pk}(f\!f)~ d(f\!f)
\end{equation}
from (\cite{Bar-etal86}). Since the statistical properties of Gaussian
fields are generally reported in terms of the peaks, an
interesting quantity is the number of peaks per cluster which can be
estimated as the ratio of the number of peaks to the number of clusters. This
ratio is shown in the lower right panel. For small filling factors
($f\!f\le 0.02$ or so corresponding to $\nu \ge 2.5$) the number of clusters
obtained numerically is unreliable and so is therefore the number of
peaks per cluster. In the range $0.02 \le f\!f \le 0.1$ ($2.5 \ge \nu \le 1.2$)
there is roughly one peak per cluster (however, the exact number
depends on the spectral index $n$).
Thus we conclude that the numerical result is roughly consistent with the
analytical calculation of Bardeen et al. (1986).
For larger filling factors the number of clusters drops
quickly due to merging into the largest cluster and the ratio of the number of
peaks to the number of cluster grows limitlessly.

\section{N-BODY MODELS}

The N-body simulations are produced by a
staggered Particle-Mesh code (Melott 1986) with $128^3$ particles on a
$128^3$ mesh
and a corresponding Nyquist wavenumber, $k_{Ny}= 64$.  The initial conditions
are
generated by the Zel'dovich approximation (\cite{Kly-Sh83}) such that
the initial power spectrum is a simple power law covering the range
$n= 3,1, 0, -1$ and $-2$.
The models are allowed to evolve gravitationally until nonlinear
effects change the slope of the power spectrum.  This change indicates that
phase correlations have developed between the originally random initial phases.

The extent of nonlinearity can be characterized by the parameter $k_{nl}$,
defined by the equation $\sigma^2_{\delta}=a^2 \int_0^{k_{nl}}P(k)\;d^3k=1$,
and in this study we evolve the simulations to values of
$k_{nl}= 64, 32, 16, 8$ and $4$ (in units of the fundamental frequency).
The value of $k_{nl}$ relates to the scale of structure
formation in real space.  For a detailed discussion of the N-body simulations
see Melott \& Shandarin (1993).

Density fields are derived from the above simulations by a cloud in cell
method whereby each particle's ``weight'' is proportionately
spread over a $2^3$ cell volume and rescaled (8:1) to produce a $64^3$
density field. This method
implies some smoothing at small scales but reduces shot noise so that further
smoothing is not needed before percolation analysis.  An ensemble family of
four realizations is produced for each combination of $n$ and $k_{nl}$ to give
assessments of the one sigma level dispersion for each percolation parameter
analyzed.

\subsection{Normalization}

It is likely that no single model studied can pretend
to explain the real universe.
We consider them to be toy models. However, if one wishes to
get a rough idea of how they may relate to the real world we provide
the following normalizations. We assume that the rms
fluctuation in number of galaxies, $\sigma_g$, is about unity within
spheres of radius $8\,h^{-1}~Mpc$, the rms mass density fluctuation $\sigma_m$
is parameterized by the ``bias factor'', $b$, such that
$\sigma_g=b\,\sigma_m$. We shall
assume that $b\approx 1$ which is an adequate assumption for these crude
estimates.
Melott and Shandarin (1993) showed that for the models in question the scale
of nonlinearity measured by the top-hat smoothing filter $R_{TH}$ is
approximately two times greater than $k_{nl}^{-1}$
calculated from the extrapolation of the linear theory
(more accurately: $R_{TH}
\approx 1.8 \,k_{nl}^{-1}$ in the $n=+1,0,-1$ models and
$R_{TH} \approx 2.8 \,k_{nl}^{-1}$ in the $n=-2$ model). Thus, identifying
every stage with the present time one can roughly estimate the
size of a mesh cell:
$l_c \approx 25,~12.6,~6.3,~3.1$, and
$1.6~~\,h^{-1}Mpc$ for $k_{nl}=64,~32,~16,~8$, and $4$ respectively.
In our models the smoothing has
been performed with a top-hat filter having a cubic rather than spherical shape
which may add an additional factor $0.6\approx (4\pi/3)^{-1/3}$ (assuming
the volumes of the filters are similar: $l_c^3 =(4\pi/3)(R_{TH}^{(s)})^3$).
Therefore, one can view each stage of the  evolution of the models as the
density distribution seen after smoothing with a top-hat filter of radius
$R_{TH}^{(s)} \approx 16,~8,~4,~2$, and $1~~\,h^{-1}~Mpc$ within
the volumes of  $(64~ l_c)^3 \approx (1600\,h^{-1}~Mpc)^3,~~
(800\,h^{-1}~Mpc)^3,~~(400\,h^{-1}~Mpc)^3,~~(200\,h^{-1}~Mpc)^3$, and
$(100\,h^{-1}~Mpc)^3$ for $k_{nl}=64,~32,~16,~8$, and $4$ respectively.
The purpose of these estimates is to give a rough idea of the range
of parameters characterizing the models, and therefore more elaborate
calculations are probably not needed.

\section{NONLINEAR GRAVITATIONAL DISTRIBUTIONS}

\subsection{Largest Cluster and Largest Void}

Figure 3 shows the largest cluster and largest void statistic for families of
four N-body models covering the complete range of initial power spectra
taken at all five stages of the evolution.
Each panel shows three curves: the largest cluster and void in the
N-body simulation, and the largest structure in the Gaussian field
having identical Fourier amplitudes with the N-body simulation. Again, in
Gaussian fields there is no statistical difference between the largest
cluster and largest void because of the symmetry of the distribution.

In Gaussian fields the percolation transition happens at
a filling factor of about $16\%$ corresponding to $\nu =\pm1$.
However, the finite size and resolution of the sample
biases the transition.
In order to avoid these effects, we obtain
the ``Gaussian'' distribution
by mixing the phases as described before.
This allows for the generation of as many Gaussian realizations with identical
amplitudes as needed to estimate the dispersion.
The Gaussian, largest structure is shown as a dotted
line in Figure 3
(hidden by the shade of the error bars) usually lying between the
solid and dashed lines.

The $n=+1$ model demonstrates the smallest difference between
the properties of the largest cluster and largest void. Both percolate
slightly, yet significantly, better than the corresponding Gaussian field.
The $n=0$ model shows that the under-dense regions percolate
similarly to the corresponding Gaussian field, but the over-dense regions
percolate better than the corresponding Gaussian field. These models exhibit
a network topology for both clusters and voids. In the $n= +3$ model
(not shown) the under-dense phase consistently percolates easier than the
over-dense phase, and the over-dense phase percolates similarly to the
Gaussian field except that it percolates easier in the last evolutionary
stage ($k_{nl}=4$).

The major feature of the nonlinear distribution is that the largest
cluster percolates easier than in the Gaussian case.
The significance of this conclusion for the largest cluster is at the
many-$\sigma$ level for most of the patterns.
Qualitatively this remains true for
other models we have
studied earlier (CDM, C+HDM (\cite{Kly-Sh93})), but
quantitatively the transitions are different. The over-dense
regions form a connected network spanning through the whole region when the
filling factor is relatively small (smaller than in the Gaussian case), and
this transition can be labeled as a shift toward
a network structure.
On the contrary, the under-dense regions may not always form a percolating
void until the filling factor of the low density phase
is significantly greater than that in the
Gaussian field. This type of transition can be labeled as a shift toward
a bubble structure.
The range of the filling factor corresponding to a sponge topology is
typically (but not necessarily) increased compared
to the Gaussian case. Thus, the above changes can also be labeled
as a shift to a sponge topology. Results (not shown) from the
percolation of
the $n=3$ models demonstrate a smooth transition from a connected topology for
both clusters and voids to the bubble topology of the $n=-2$ models.  In
addition, the $n=3$ case supports our claim of the universality of a
filamentary nature for the mass distributions.
However, the major point is not how to label a structure but rather
to show that in a general case the two shifts are independent of each other
and carry independent information about the structure. Therefore, combining
them into one parameter (like $W_{\nu}=\nu_+ - \nu_-$ mentioned above)
results in the loss of information.
Similar to Gaussian fields
at small filling factors, the largest structure must be negligible in
sufficiently large samples to insure a fair sample.

\subsection{Density Contrast of the Largest Cluster}

The significance of the percolation transition in the nonlinear gravitational
distributions sometimes is challenged on the grounds of the low density
threshold of the percolation onset.
However, these numbers may be misleading.
We believe that the mean density is more relevant. Let us imagine that we
have two identical density contours with $\delta =0.5$ in two dimensions.
Within one of them there are a few peaks the highest of which is
$\delta\approx 1$
and within the other one there are few peaks of order $\delta \approx 10$.
Now the question is which of these two contours will be picked up by the eye?
We believe that obviously the latter is more noticeable. In the linear regime
both the mean density and the density threshold are close,
but in the nonlinear regime they are very different. Figure 4 shows the mean
density of the largest cluster as a function of its volume given in units
of the filling factor. For comparison, the density contrast is plotted as a
thin line and is markedly below the mean density in every case. An interesting
feature of this figure is the stability
of the mean density of the largest cluster through the percolation
transition. The other important aspect of the figure is the value of the mean
density at percolation. At percolation the mean density of
the largest cluster is well above the average density of the distributions
and understandably increases with evolution and decreasing
spectral index. The double-valued nature of the function for the early stages
of some models is an effect of a finite system explained above (section 2.2).

\subsection{Number of Clusters and Voids}

Another statistic characterizing the density distribution is the
total number of clusters and voids as a function of the filling factor.
As mentioned before this statistic is sensitive to the slope of the spectrum
in the Gaussian fields. The number of clusters and voids is normalized
to the Nyquist volume (8 mesh cells) as was done for Gaussian distributions.
Frequency distribution studies establish that the smallest structures
(both clusters and voids)
dominate in the total number of structures for all spectra studied.
Figure 5 shows this statistic for all models and all stages (except $n= 3$).
Three curves are plotted for each pattern.
The total number of clusters and the
total number of voids in the nonlinear distributions, and the total number
of clusters (or voids) in corresponding Gaussian fields
after randomization of phases. Typically the number of both clusters and
voids in the parent N-body simulation is less than that in the corresponding
nonlinear Gaussian field.
Thus, the phase correlation due to nonlinear gravitational effects reduces
the number of both over-dense and under-dense structures. The smaller the
spectral index, $n$, the stronger the effect.
The number of voids is consistently greater than the number of
clusters in the nonlinear distributions if $n\le0$. On the other
hand the number of clusters is greater in the $n=+1$ model
\footnote{The last two stages ($k_{nl}=8$ and $4$)
in the $n=+1$ model and the very last stage ($k_{nl}=4$) in the $n=0$
model suffer from
discreteness. The voids are completely empty and percolate at such
low density thresholds that the onset of percolation cannot be reliably
calculated in the N-body simulations in question.}.  Another important trend is
that at the later stages the differences between models become
weaker compared to the beginning stages. With evolution, the models tend toward
an estimated slope of
$n\approx -1$ in the nonlinear regime which is in very
good agreement with the direct measurement of the spectra (\cite{Mel-Sh93}).

Using the right middle panel of Figure 2 as a calibration, one can estimate
the slope of the spectra for the nonlinear distributions by extrapolating
between the maximum values from the pure power law, Gaussian fields.
The maximum values from both types of Gaussian fields are arranged in
Table I with the
estimated slopes in parentheses. Although the table is generally consistent
with the trends described above, two anomalies
are apparent in the results.  First, the maximum value
for the case $n=-1$, $k_{nl}=64$ is considerably below the corresponding
Gaussian value; and second, the models $n=-1$ and $n=-2$ do not
monotonically approach a limiting slope of -1. Both of these peculiarities
have a single explanation. Close inspection of the evolved power spectra
for these models (see Melott \& Shandarin 1993) shows that the same conditions
apply to the slopes of the power spectrum if measured at $k=32$. The Nyquist
frequency of the Fourier transform used to produce the nonlinear Gaussian
fields is $k_{Nq}=32$. This suggests that our randomization procedure is
sensitive to the slope of the power spectrum of the parent field at the
maximum value of $k$ used in the Fourier transform.  In these cases, the
density fields were $64^3$ fields with maximum $k$ values equal to the Nyquist
wavenumber, but other cutoff values, $k_c \le k_{Nq}$, could be enforced.
Further study is needed to make the dependence of the
slope on the cutoff ($k_c$) of the randomization procedure a practical
technique to measure the effective slope of spectra. The models
$n=1$ and $n=0$ do not exhibit the anomalies explained above because their
power spectra correspond to the trends at the Nyquist frequency.

\section{DISCUSSION}

The quantitative topology proved to be a useful measure for studying the
large-scale structure in the universe. We use a percolation technique to study
the mass distribution in real space obtained from N-body simulations of power
law, initial spectra: $P(k) \propto k^n$ with $n=+3,~+1,0,~-1,$ and $-2$ in
the $\Omega=1$ universe. Five stages corresponding to the scale of
nonlinearity at $k_{nl}=64,~32,~16,~8,$ and $4$ are analyzed. Each model was
run with four different sets of random numbers to create ensemble families so
that the dispersion of statistical parameters could be estimated.

We discuss in detail only two robust characteristics calculated in
percolation analysis: 1) the normalized volume of the largest cluster
(or void), and 2) the total number of clusters (or voids) as
a function of the filling factor.  It is important to
stress that the statistical properties of clusters and voids are
identical in Gaussian fields, but they are significantly different in the
nonlinear distributions drawn from the N-body simulations.
Thus both statistics are sensitive to the non-Gaussianity of a field.  These
two functions can be very efficient discriminators of the models.  All of
the density distributions studied here have unique sets of the three functions
(the filling factor, the largest structures, and the number of structures)
on a multi-sigma level.

Combining the results that the most numerous clusters are the size of
the cutoff scale, and that the nonlinear distributions percolate better
than the Gaussian fields, we
conclude that the small clusters are arranged in accordance with an underlying
network structure. A similar conclusion was derived from the adhesion
approximation (\cite{Kof-etal92}),
and the truncated Zel'dovich approximation (\cite{Col-Mel-Sh93};
\cite{Pau-Mel95}).
The largest cluster as well as the largest void
statistics indicate the percolation thresholds which
are associated with a change in the topology of the distributions. The onset
of percolation in the over-dense phase marks the transformation from
a clumpy (meat-ball) to
a network (sponge) topology, and the onset of percolation in the under-dense
phase marks the transition from a bubble to a network topology.
Comparing the simulations with the Gaussian fields we conclude:

1) For all models with $n\le +3$
over-dense regions percolate better than the corresponding Gaussian fields
showing a shift toward a network structure.
The later the stage the more significant the shift.
Since the models are approximately scale free
the later stages can also be interpreted as earlier stages of the same
distribution seen with better
resolution. Thus, we conclude that all power law models with $n\le +3$ show
a network structure in the density distribution at the nonlinear stage
if seen with sufficient resolution. Any differences are only quantitative, but
highly significant. The smaller the spectral index the
stronger the shift.
{}From this, we conclude that all realistic cosmological
models (with non-power law, initial spectra like the CDM family)
can be characterized as filamentary
or network structures as far as the mass density in real space is concerned.
At early stages of the CDM or C+HDM models when the effective slope at the
scale of nonlinearity is small
enough ($n_{eff}< -1$), the density distribution is also of a bubble type.

2) The percolation transition happens at relatively low density
contrasts. The larger the value of $n$, the smaller the density contrast at the
percolation threshold (compare Figures 3 and 4).
However, the density threshold is a misleading parameter.
We propose the mean density of the largest cluster
to characterize the prominence of the transition. The value of the mean
density of the largest structure at the percolation transition indicates
a definite nonlinear character for the largest cluster.

3) Under-dense regions percolate at lower filling factors than their Gaussian
counterparts for $n= +3$ and $+1$, and at
considerably higher filling factors for
$n= -2$. There is a smooth and consistent trend for under-dense regions
to percolate with greater difficulty as the spectral index decreases or with
evolution of a simulation. This trend represents a shift towards a bubble
topology for models with $n\le -1$ indicated by the under-dense regions
percolating at higher filling factor values than the corresponding Gaussian
fields while the over-dense regions percolate at lower filling factors
than the corresponding Gaussian fields.
There is a noteworthy quantitative difference between the models
$n=-1$ and $n=-2$.

4) If the density threshold is selected so that the
over-dense and under-dense phases each occupy $50\%$ of the volume,
then all models show that the largest cluster and void occupy almost all of the
space. All smaller structures combined occupy at most a few percent of the
volume for all the power law models studied. Thus, the distribution can be
labeled as a sponge topology. Table II
gives the fractions of the total volume for all clusters
and voids (except the largest structures) when the filling factor is 0.5, and
the corresponding values for the Gaussian fields. Despite the smallness of the
numbers they are highly significant.
This explains why Weinberg, Gott \& Melott (1987) found that the nature of the
interlocking, equal-volume, over-dense and under-dense regions
of a random field was a ``sponge topology'' where both the under-dense and
over-dense regions are equivalent.

5) Our results show that all approximations for a nonlinear, gravitational
instability based on nonlinear transformations of initial Gaussian density
fields (e.g. the lognormal model) make incorrect predictions of the topology
since they preserve the similarity in percolation properties of the over-dense
and under-dense phases to that of the Gaussian fields.

We acknowledge the support of NSF grant AST-9021414, NASA grant NAGW-3832
and the University of Kansas GRF-95 grant. We thank Adrian Melott for
discussions and for supplying the simulations which were all produced at the
National Center for Supercomputing Applications. We also thank L. Kofman,
V. Sahni, and B. Sathyaprakash for discussions and comments.

\vfill
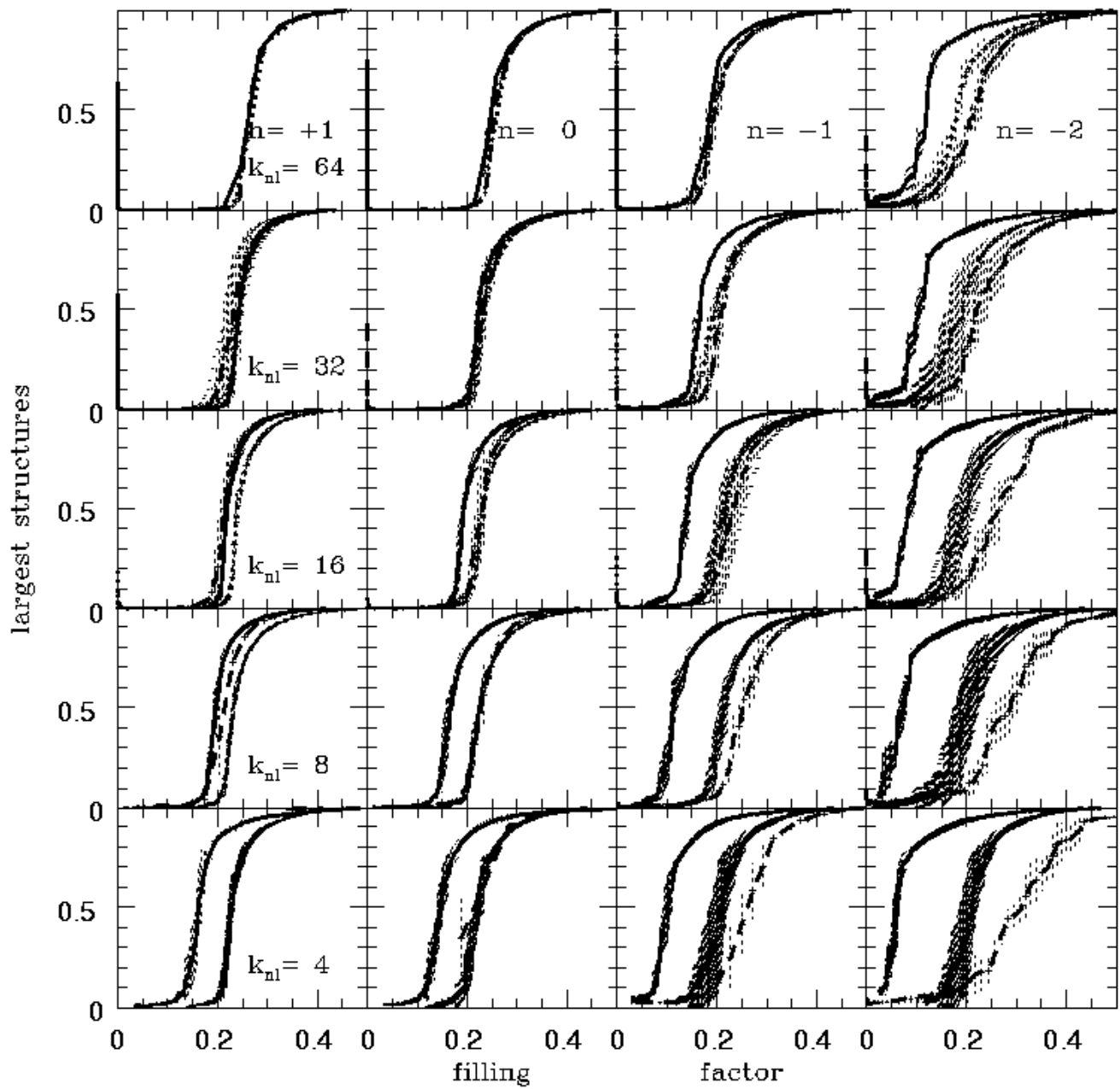
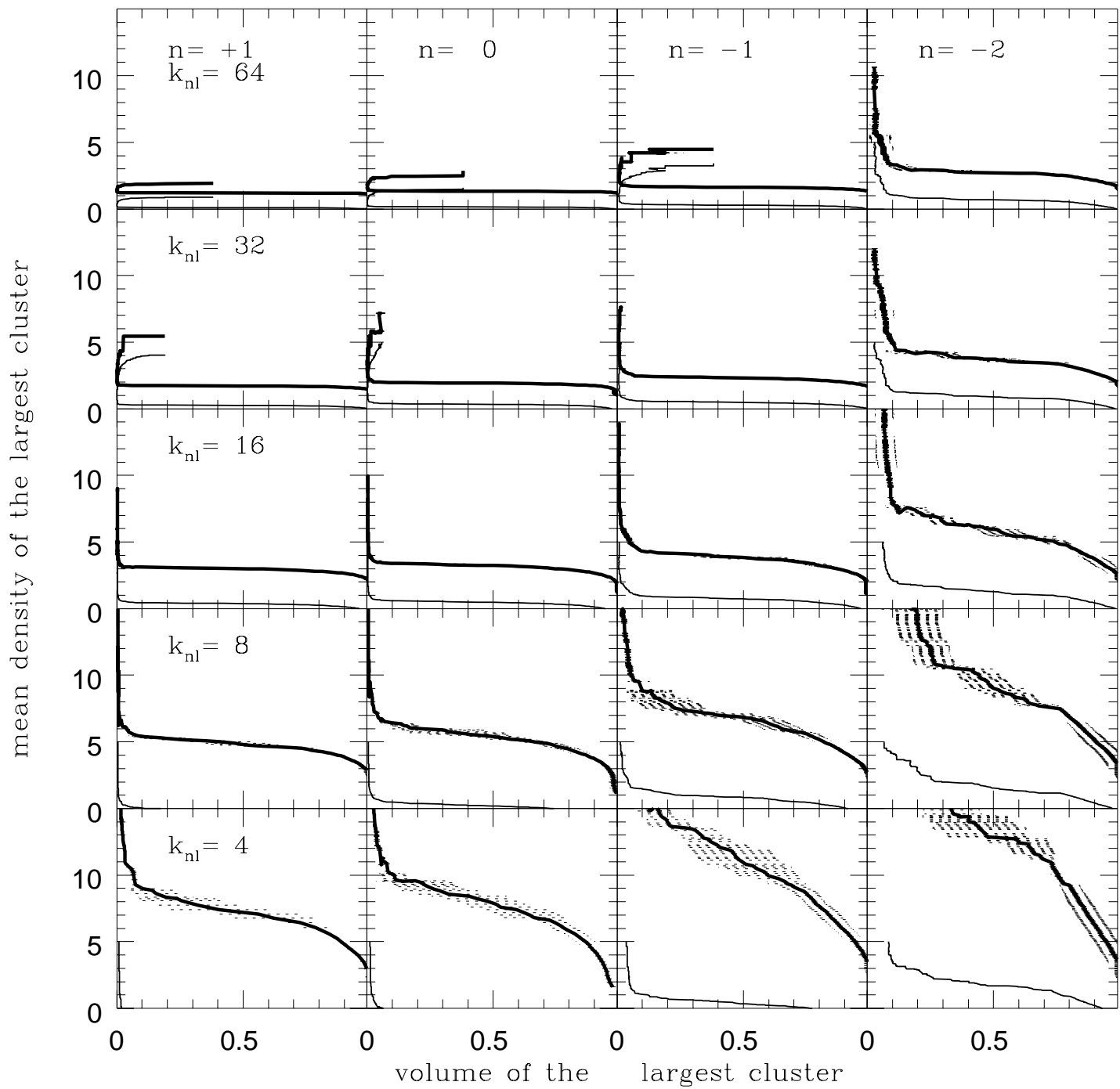
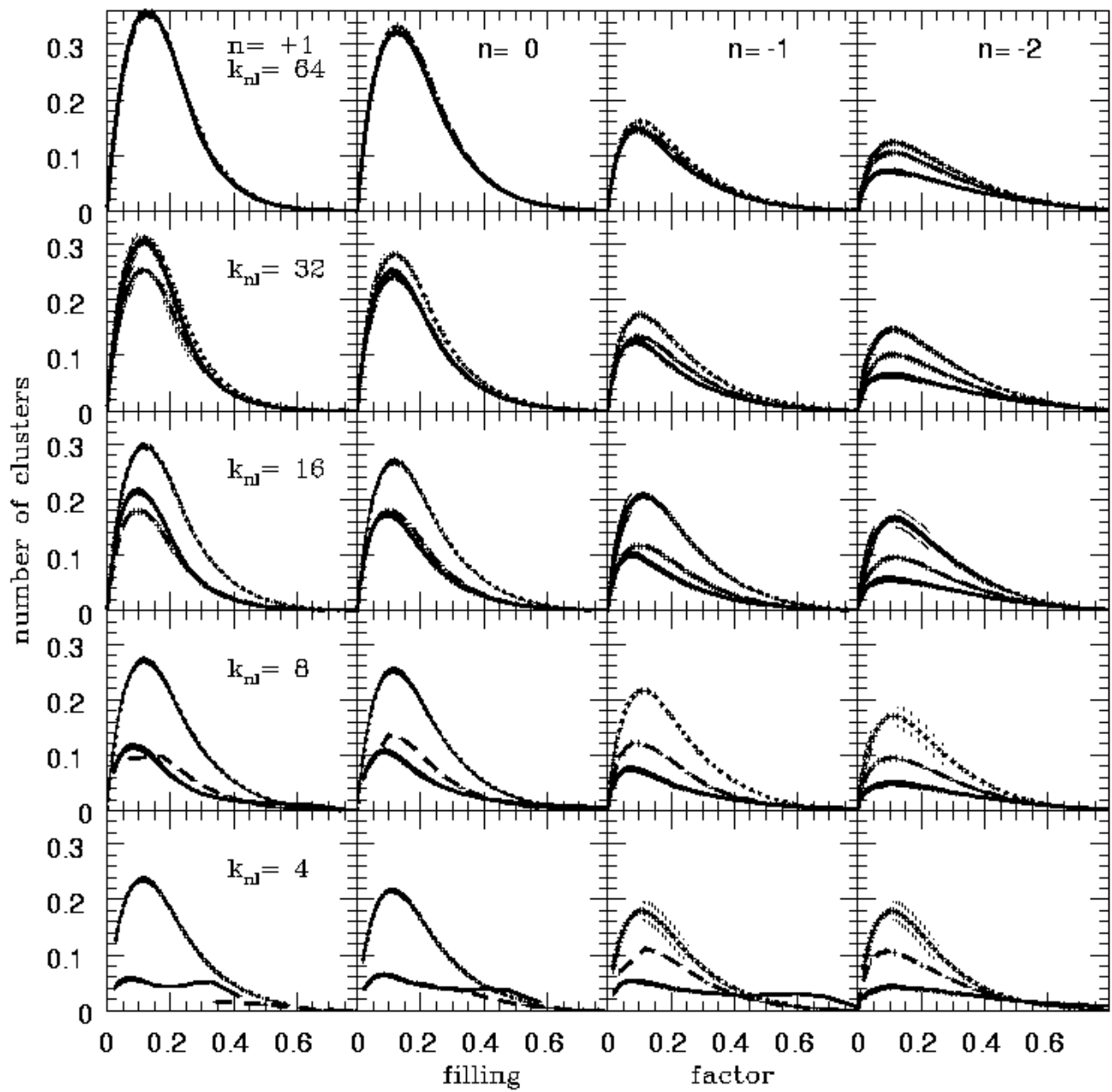
\end{document}